\documentclass[a4paper,twoside]{article}
\usepackage[T1]{fontenc}
\usepackage{amssymb}
\usepackage[dvips]{graphicx}

\newcommand{\C}{\mathbb{C}}
\newcommand{\cH}{\mathcal{H}}
\newcommand{\tens}{\otimes}

\title{Schr\"odinger's cat and the clock:\\ Lessons for quantum
  gravity}
\author{Robert Oeckl\footnote{email: oeckl@cpt.univ-mrs.fr}\\ \\
Centre de Physique Th\'eorique,\\
CNRS Luminy,
13288 Marseille, France}
\date{CPT-2003/P.4542\\ September 19, 2003}

\begin{document}
\maketitle

\begin{abstract}
I review basic principles of the quantum mechanical measurement
process in view of their implications
for a quantum theory of general relativity.
It turns
out that a clock as an external classical device associated with the
observer plays an essential role. This leads me to postulate a
``principle of the integrity of the observer''. It essentially requires
the observer to be part of a classical domain connected throughout the
measurement process.
Mathematically this naturally leads to a formulation of
quantum mechanics as a kind of topological quantum field
theory. Significantly, quantities with a direct interpretation in
terms of a measurement process are associated
only with amplitudes for connected boundaries of compact regions of
space-time. I discuss some implications of my proposal such as in-out
duality for states, delocalization of the ``collapse of the
wave function'' and locality of the description. Differences to
existing approaches to quantum gravity are also highlighted.
\end{abstract}

\section{Introduction}

In spite of many decades of research a quantum mechanical theory of general
relativity still seems out of reach. The opinions as to why this is so
are divided, ranging from blaming purely technical difficulties to
claiming a fundamental incompatibility between quantum mechanics and
general relativity. Although the latter point of view might seem
extreme, the persistent failure of reconciling the two frameworks
justifies at least a review of fundamental principles. I wish to
contribute to such a review by reexamining the measurement process in
the context of a quantum description of space-time.

Conventionally, the problem of quantum mechanical measurement is
treated as something
that can be considered in the context of a classical (and even
non-special-relativistic) space-time.
This is what the formalism of quantum mechanics is based on. It is
then assumed that a quantization of space and time is merely a second
step which can be performed within the formalism thus set up.

In contrast to this point of view I emphasize here the desired quantum
nature of space and time in the formalization of the measurement process
itself.
My guiding principle in doing so does not consist of introducing any
new postulates or of
proposing an alternative to quantum mechanics. Rather, it consists of
``taking
serious'' basic principles of quantum mechanics while not
necessarily endorsing every aspect of its standard formalism.
Indeed, it is precisely the
application of principles of quantum mechanics itself to space-time
which prompts me to argue for a modification (or rather
generalization) of the standard formalism.

\section{Schr\"odinger's cat revisited}

\begin{figure}
\begin{center}
\includegraphics{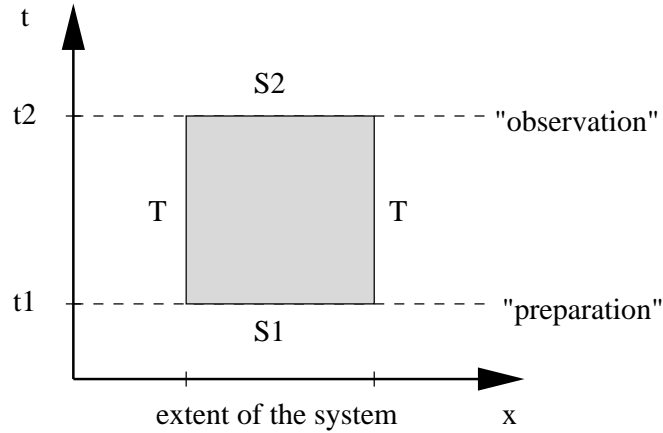}
\caption{The space-time diagram of the measurement process. Only one
  spatial dimension is shown. The world-line of the system between
  preparation at time $t_1$ and observation at time $t_2$ is the
  shaded area.}
\label{fig:stdiag}
\end{center}
\end{figure}

As an illustration of the prototypical quantum mechanical measurement
process I recall the thought experiment with Schr\"odinger's cat
\cite{Sch:katze}.
Assume I take a box into which I put a living cat and a
radioactive isotope connected to a mechanism that will kill the cat if
a decay is detected. I close the lid (that is, I isolate the system
from its environment) and wait for a certain time
$t$. Then I open the lid to see whether the cat is dead or still
alive. I find (repeating the experiment many times) that the cat will
still be alive with a certain probability $p(t)$, but I cannot predict
the outcome of any \emph{single} experiment.
Quantum mechanics forbids me to assume that the
cat is either definitely dead or definitely alive while I do not
look into the box.
In other words, it disallows me to assume a definite classical
evolution to take place inside the closed box.
Indeed, as decades of experimental evidence in quantum mechanics have
shown, such an assumption would lead to contradictions.\footnote{Of
 course this cannot really be said for
 a macroscopic cat. But this is not the point here; the reader might
 imagine instead a ``microscopic cat''.}
Nevertheless, quantum mechanics allows me to exactly calculate the
probability $p(t)$.

Generally, the measurement process involves a quantum
domain and a classical domain. The system on which the
measurement is performed (here the interior of the box) is part of the
quantum domain while the observer is part of the classical domain
(here the surroundings). In the quantum domain no definite classical
evolution takes place and it does not make sense there to ascribe
classical states to the system.

The conventional mathematical description of the experiment is as
follows (see Figure~\ref{fig:stdiag} for a space-time
diagram). Associated with the system is a
Hilbert space $\cH$ of states. At the time $t_1$ I prepare the system in
a state $\psi\in \cH$ (I set up isotope, machine and cat). Then I
isolate the system (I close the lid). I let the system evolve for a
time $\Delta t$, which is described by a unitary operator $U(\Delta t)$
acting on $\cH$. At time $t_2$ I perform the observation (I open the
lid). That is, I can ask whether the system is in a state $\eta\in \cH^*$
(e.g.\ if the cat is still alive). The probability $p$ that this is the
case is the modulus square of the corresponding transition amplitude:
\[
p(\Delta t)=|\langle \eta| U(\Delta t)| \psi\rangle|^2 .
\]

\section{The cat and the clock}
So far I have implicitly assumed that space and time provide a fixed
classical background structure. Now I want to treat them as quantum
mechanical entities as well. If I take serious the principle that I
can know nothing classical about the interior of the box while I do
not look inside then this must also extend to space and time. In
particular, I must not assume any definite (classical) passage of time
inside the box. 
Outside the box on the other hand time remains a classical entity as
part of the classical domain of my observations.

To examine the implications of this I need to pay more attention to
how the notion of time enters into my measurement process. I
thus refine the description of the experiment as follows: Next to the
box I put a clock. After closing the lid I continuously watch
the clock until I find that the time $\Delta t$ has elapsed. Then I proceed
to open the lid and look inside the box. Of course, the outcome of the
experiment is as before, I observe the same probability $p$ of the cat
still being alive as a function of $\Delta t$.

But how can this be? How does the system ``know'' about the time
$\Delta t$
elapsed on my clock when I cannot assume any definite evolution of
time inside the closed box?

The sensible answer seems to be that the system \emph{does} stay in
contact with
its environment while the lid is closed. More precisely, it
stays in contact with the space-time as classically
experienced by me as the observer. The
information about the space-time structure surrounding the box is a
\emph{boundary condition} to the experiment. It encodes in particular the
elapsed time $\Delta t$ on the clock. \emph{This boundary condition must
be regarded an integral part of the quantum mechanical measurement
process.}

Of course, the focus on the time variable just serves
to emphasize my point. What is relevant are both, time and space.
Indeed, it is important that I do not move around the box after
closing the lid. For example, I do not allow the box to be taken away
to (say) Alpha Centauri C and brought back so that the time dilatation
would alter the function $p(\Delta t)$ and thus the outcome of my
measurements. 
Even with space-time behaving completely classically the
inclusion of this boundary condition into the measurement process can be
useful. With a quantum mechanical space-time it becomes inevitable.

\section{Integrity of the observer}
On the conceptual level I can view the above argument also as
stating something about the observer. Let me call this the \emph{principle
of the integrity of the observer}. This means that the whole
measurement process (including preparation and observation proper)
pertains to one connected classical domain in which the observer
describes reality. In the above thought experiment this connectedness
is manifest in the clock and in me as the observer watching the clock
while the box is closed.

A measurement distributed over several disconnected
classical domains does not make sense. An observer in one of them
would have to relate to other ones either through classical channels
of communication or via interaction through the quantum domain. In the
first case the classical domains would be connected into just one
classical domain. In the second case the other classical domains
effectively become part of the quantum domain.
Applied to the prototypical experiment described above this means that
it does not make sense for the observer to consider preparation and
observation as disconnected interactions between classical and quantum
domains. To the contrary, to relate the two it is essential that the
observer has a classical existence in between (with a classical time
duration $\Delta t$).

\section{Criticism: The clock in the box}
Before proceeding I would like to meet one possible criticism of my
interpretation of the thought experiment with Schr\"odinger's cat.
I have argued that the elapsed time $\Delta t$ on my clock next to the
box is
something about which the system inside the box cannot ``know'' anything
in the conventional interpretation of the experiment. But what
if I put the clock \emph{inside} the box? Upon opening the box I could
correlate this internal time $\Delta t_i$ to the probability of finding the
cat dead or alive. I would obtain the same probability distribution as
above. It seems that this way of looking at the experiment
does not require any kind of contact between system and outside world
while the lid of the box is closed.

But I have a problem now. After closing the lid, when shall I open it
again to look for the cat (and the clock)? As the clock is inside the
box I cannot look at it while I wait. (Indeed
this is essential for it to be part of the quantum domain.)
Shall I think of the time of opening the box as somehow ``random''?
But how? Perhaps I should view this as somebody else preparing the
experiment for me and sending me a closed box. This I open and
determine both $\Delta t_i$ and the state of the cat. But as there is
no such thing as an a priori random distribution of the $\Delta t_i$
this might not
tell me anything. For example, in the repetitions that I arrange it
could be that the 
$\Delta t_i$ is always astronomically large and the cat always dead.

The point is that the modification \emph{fundamentally changes} the
experiment. It does \emph{not} provide me with the same information as the
original experiment. The fact that I can predetermine the elapsed time
$\Delta t$
is an essential part of the original experiment. The modified
experiment thus corresponds to a different measurement process.
Apart from that, it is not even
clear whether this modified measurement process can be made sense of without
introducing some classical connection between preparation (even if
this is done by ``somebody else'') and
observation through the back door. For example, what distribution of
times $\Delta t_i$ am I supposed to observe opening many boxes?

A different strategy would be to put a clock outside the box as well.
But then, how do I explain that the external and internal clocks show
the same elapsed time (up to quantum fluctuations)? This leads me back
to my original argument.

\section{Encoding the boundary condition}
Let me propose a formalization of the consequences of the thought
experiment.
To this end I shall assume that I can strictly identify classical and
quantum domain with corresponding regions of space-time.
In the present experiment the quantum domain is thus the world 4-volume of
the box in the time interval $[t_1,t_2]$ (the shaded
area in Figure~\ref{fig:stdiag}).  The classical domain
is everything outside. (This assumption will turn out to be
stronger than required.)

Now, how exactly is the ambient classical
space-time a boundary condition to the experiment?
According to quantum
mechanics the interaction between observer and system (preparation
and observation) should take place at the interface between the
classical and quantum domains. 
Thus, the relevant spatio-temporal information should reside
in the metric space-time field on the three-dimensional boundary that
separates the two. 
This three-dimensional
connected boundary $B$ consists of three parts (see
Figure~\ref{fig:stdiag}):
the space-like boundary $S_1$ consisting of the inside of the box at the
time $t_1$ of preparation, the
time-like boundary $T$ of the spatial boundary of the box while
waiting for the time
$\Delta t$ to elapse and again the space-like boundary $S_2$ of the
inside of the box at the time $t_2$ of observation.

To what extent a metric on a connected surface such as $B$ determines
or over-determines a solution of the Einstein equations inside is a
difficult initial value problem. Due to the similarity with the
``thick sandwich'' problem \cite{Whe:gmfinal} I shall assume that the
intrinsic metric is sufficient. However, the exact validity of such an
assumption is probably not a crucial ingredient for a quantum
theory of general relativity.

\section{Quantum general relativity as a TQFT}
To calculate a transition amplitude I
need thus three pieces of information: The initial state $\psi$ on
$S_1$,
the final state $\eta$ on $S_2$ and the intrinsic metric $g$ on the
whole of $B=S_1\cup T\cup S_2$. This suggests a mathematical
description as follows: Associated with $B$ is a state space $\cH_B$
and I can think of $(\psi,\eta,g)$ as determining an element in this
space.
(In a truly quantum description of space and time I should think of
$g$ really as a quantum state ``peaked'' at a classical metric
rather than a classical metric itself.)
The amplitude for such a state is given by a map
$\rho:\mathcal{H}_B\to \C$. The associated probability (density) $p$ is
as usual the modulus square of the amplitude, i.e.\
\[
 p=|\rho(\psi,\eta,g)|^2 .
\]

To recover the conventional mathematical
description of the experiment the state space can be split into a tensor
product corresponding to the boundary components
$\cH_B=\cH_{S_1}\tens\cH_{T}\tens\cH_{S_2}$. Correspondingly, I label
the metric living on the different components by $g_1,g_T,g_2$. Then
I should recover $\cH=(\cH_{S_1}|g_1)$, i.e.\ the state
space $\cH$ is the space of states in $\cH_{S_1}$ which are partly fixed
to $g_1$ (namely in their metric information), correspondingly
$\cH^*=(\cH_{S_2}|g_2)$. I call $\cH$ and $\cH^*$ \emph{reduced state
  spaces}.
The amplitude is then equal to
\[
\rho(\psi,\eta,g)=\langle
\eta_{g_2}| U(g_T) | \psi_{g_1}\rangle .
\]
The indices on the states
indicate that they live in the reduced state spaces with fixed metric
and the argument of $U$ that it depends (apart from the state spaces
$\cH$ and $\cH^*$) on the metric $g_T$. Note that in particular, $g_T$
contains the information about the time duration $\Delta t$.

The mathematical structure of the formalization I have given above is
essentially that of a topological quantum field theory (TQFT)
\cite{Ati:tqft}. Let me briefly recall what this means in the present
context. The basic setting is that of topological (or rather
differentiable) manifolds in dimension four.
Associated to any 3-boundary $B$ of a
4-manifold $M$ is a vector space $\cH_B$ of states.
Associated to $M$ is a map
$\rho_M^{B,\emptyset}:\cH_B\to \C$. If $B$ is the union of components
$B=B_1\cup B_2$
the space $\cH_B$ is the tensor product of spaces associated with the
components $\cH_B=\cH_{B_1}\tens \cH_{B_2}$.\footnote{Usually one
allows only closed boundaries. The above decomposition $B=S_1\cup
T\cup S_2$ would require boundaries with boundaries as well. This
requires  what is
also called a ``TQFT with corners''.}
Besides considering
boundaries as ``in'' I can also regard them as ``out''
whereby the corresponding space is replaced with its dual and appears
in the codomain of $\rho$ instead of its domain. For example,
regarding $B_1$ as ``in'' and $B_2$ as ``out'' I would
have $\rho_M^{B_1,B_2}:\cH_{B_1}\to \cH_{B_2}^*$.

\begin{figure}
\begin{center}
\includegraphics{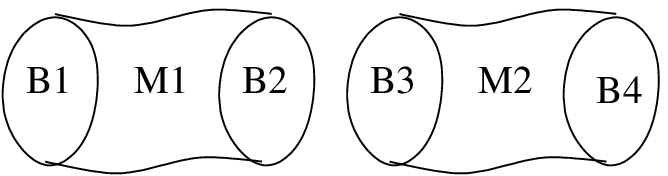}\\
\vspace{1cm}
\includegraphics{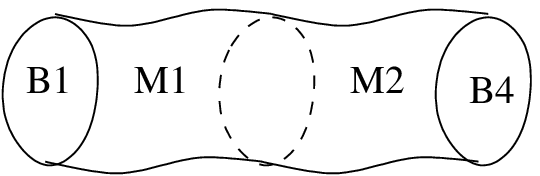}
\caption{Gluing two manifolds along a common boundary corresponds to
  composition in TQFT.}
\label{fig:comp}
\end{center}
\end{figure}

A crucial property of a TQFT is the composition property: The gluing
of two 4-manifolds at a common boundary corresponds to the composition
of the respective maps. More precisely, consider a 4-manifold $M_1$ with
boundaries $B_1$ and $B_2$ and a 4-manifold $M_2$ with boundaries
$B_3$ and $B_4$. Then I have maps $\rho_{M_1}^{B_1,B_2}:\cH_{B_1}\to
\cH_{B_2}^*$
and $\rho_{M_2}^{B_3,B_4}:\cH_{B_3}\to \cH_{B_4}^*$. Assume that $B_2$
and $B_3$ are
mirror images so that $M_1$ and $M_2$ can be glued together (see
Figure~\ref{fig:comp}). This also
implies $\cH_{B_2}^*=\cH_{B_3}$. The composition property then states
for $\rho_{M_1\cup M_2}^{B_1,B_4}:\cH_{B_1}\to \cH_{B_4}^*$ associated
with the union $M_1\cup M_2$ that
\[
 \rho_{M_1\cup
 M_2}^{B_1,B_4}=\rho_{M_2}^{B_3,B_4}\circ\rho_{M_1}^{B_1,B_2} .
\]
If I consider the special case of boundaries that are
equal-time-slices in Euclidean or Minkowski space I recover the usual
formulation of quantum mechanics. The composition property corresponds then
to the summation over a complete set of intermediate states.

However, my point is that I allow more general
boundaries (and here I go beyond the use that is usually made of TQFT
in physical contexts). In particular, boundaries
might have time-like components and I may glue along such
boundaries. Rather than a composition of time-evolutions this would
be a composition in space. Physically, this might for example
correspond to the formation of a composite system out of separate
systems. Since this is really a ``quantum'' composition in the same
sense as the composition of time-evolutions is, the consistency of
this operation is ensured.

\section{In-out duality}
An implication of allowing rather arbitrary boundaries is that the
distinction between ``in'' and ``out'' states of quantum
mechanics becomes
arbitrary, in turn blurring the distinction between preparation and
observation proper in the measurement. Even the ``in'' and ``out''
notions of TQFT become inadequate. More precisely, they become purely
technical notions that can essentially be turned around at will
(by dualization, see above). The \emph{physical} notions of ``in'' (as
preparation) or ``out'' (as observation) become necessarily
disentangled from this technical one.
For example, in the above thought experiment
certain data associated with the boundary component $S_2$ was
considered ``out'' (the reduced state $\eta$) while certain other
data on the same $S_2$ (the metric or quantum state of space-time
$g_2$) was considered ``in''.

So how is this separation into physical ``in'' and ``out'' encoded?
The answer is that there is no need for it.
This is not really a surprise. Indeed, quantum field theory has been
teaching us this lesson for a long time. It is manifest in a
remarkable feature of the LSZ reduction \cite{LSZ:reduction}.
Consider the time ordered correlation function (in momentum
space) of $n$ fields,
\[
\langle 0| T \phi(p_1) \cdots \phi(p_n)|0\rangle .
\]
Its modulus square is a probability density -- and in several ways.
Given incoming particles
with momenta $p_1, \dots, p_k$ it is essentially the probability
density for observing outgoing particles with momenta $p_{k+1}, \dots,
p_n$.\footnote{I am simplifying slightly here by leaving out operators of
the type $(\Box+m^2)$ that have to be applied to the $n$-point
function. However, this is not relevant to the discussion.} 
How many particles I regard as incoming (i.e.\ which value
I take for $k$) is arbitrary. For each choice the right answer is
given by the very same quantity. How I split up the ``state'' into
prepared part (corresponding to $\psi$ above) and observed part
($\eta$ above) is arbitrary.
Note though that exchanging ``in'' and ``out'' states requires to
exchange positive with negative energy. But this fits the
associated orientation reversal in the time direction of the TQFT
description.

In the same sense the function $\rho$ is to be regarded as giving an
amplitude for a state $\psi$. Whether a part of this state is to be
considered as prepared or as observed does not alter the
associated probability density. It is rather to be viewed as an
ingredient of the experimental circumstances. This appears
to be a rather strong postulate in general but I hope to have made it
plausible through well established physics.

The removal of the a priori distinction between preparation and
observation proper might also be seen as having implications for the
interpretation of quantum mechanics. In the conventional picture I can
think that I prepare the system in the state $\psi$ at time $t_1$,
after which it evolves deterministically to the state
$U(\Delta t)\psi$. Then I perform the observation at time $t_2$ and the
wave function ``collapses''. This description of course no
longer makes sense in the proposed formulation. There the boundary is
connected and it seems rather far-fetched to associate a ``collapse''
with any particular piece of this boundary (especially if it has a
quite arbitrary shape). Instead one could still talk about a
``collapse'' but this would have to be associated with the boundary as
a whole. In particular, it is no longer localized in time and thus
cannot have the usual connotation of the ``instantaneous disruption''
of a deterministic evolution.

\section{Connectedness of the boundary}
Above I have argued how the thought experiment naturally leads to a
TQFT description of quantum mechanics.
Significantly, the principle of the integrity of the observer implies
the \emph{connectedness} of the boundary at the interface between
classical and quantum domain. That is, a TQFT amplitude can only have a
direct interpretation in terms of a quantum mechanical measurement
process (that involves a quantum treatment of space-time) if it is
associated with a connected 3-boundary. 

One might object that ordinary quantum mechanics gets along very well
with disconnected boundaries. But this I would argue, is due rather to
simplifications (especially due to the fixed space-time background)
than fundamental reasons. A typical system of interest
has a finite extent. Outside this extent nothing
relevant happens that requires really a quantum mechanical
treatment. Comparing this to Figure~\ref{fig:stdiag}, nothing
interesting happens at the boundary $T$ and it might be neglected and
conveniently pushed to infinity. The situation becomes different
however, when space-time is no longer regarded as a fixed background
but treated quantum mechanically as well. As I have argued, the
boundary $T$ then plays an essential role.

Note that this argument also remains valid if the system is infinitely
extended. The crucial point is that the observer remains excluded so
that there is a boundary between him and the system. Pushing this
further we might even ``invert'' the picture of
Figure~\ref{fig:stdiag} and consider the observer's world line as
surrounded by a boundary outside of which ``the quantum mechanics
happens''. I will not pursue this point of view here though.

\begin{figure}
\begin{center}
\begin{tabular}{cp{1cm}c}
\includegraphics{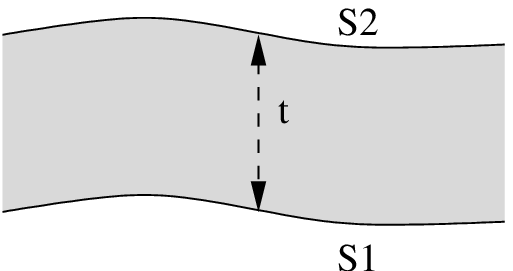} &&
\includegraphics{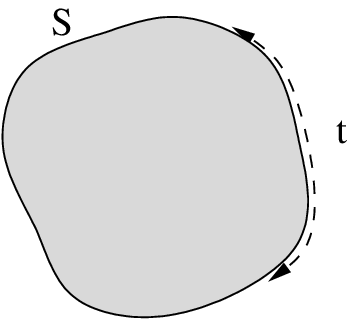}\\
\\
(a) && (b)
\end{tabular}
\caption{(a) The more traditional approach to quantum gravity employs
  space-like boundaries $S_1$ and $S_2$. (b) The advocated approach
  employs connected boundaries $S$ of compact regions of space-time.}
\label{fig:boundaries}
\end{center}
\end{figure}

The connectedness of the boundary is
rather significant for the interpretation of theories of
quantum gravity and quantum cosmology.
Let me compare this to the more traditional point of view that is
often adopted in approaches to quantum gravity (e.g.\ in the
Wheeler--DeWitt approach \cite{Dew:qgrav1,Whe:sspace}, in Euclidean
quantum gravity \cite{Haw:pathintqg} and also in loop quantum gravity
\cite{Rov:lqg}). 
Consider two space-like boundaries (say Cauchy surfaces) $S_1$ and
$S_2$ which are
closed and extend ``to infinity'' in the universe (see
Figure~\ref{fig:boundaries}.a). One then considers transition
amplitudes between quantum states of the metric on $S_1$ and on
$S_2$. The question how a time duration $\Delta t$ (along some path)
between an event on $S_1$ and an event on $S_2$ is to be encoded is
then answered as follows:\footnote{I would like to thank Carlo Rovelli
for elucidating me on this crucial point. Note also that this argument
does not work for the ``degenerate case'' of Minkowski space, as is
rather obvious.} Given a solution of Einstein's
equations consider the two non-intersecting spatial hypersurfaces $S_1$
and $S_2$. Then, generically, it is conjectured that this solution can
be reconstructed (up to diffeomorphism) given the intrinsic 
metrics on $S_1$ and $S_2$. (This is the ``thick sandwich'' problem
\cite{Whe:gmfinal}.) This implies that such intrinsic metrics
contain the information about the time difference in the above
sense. At least for quasi-classical states in a suitable sense it
should be appropriate to talk about time durations $\Delta t$ (with
some uncertainty) between ``initial'' and ``final'' state.

Nevertheless, this approach has the disadvantage that it cannot
be directly related to a measurement process of the type considered
above. What I called the
principle of the integrity of the observer is violated. To remedy this
I would presumably have to fix some spatial region (where I as
the observer live) and its world-line to be classical. But this would
essentially amount to introducing extra boundary components that
connect $S_1$ and $S_2$, thus introducing a connected boundary through
the back door.
In the proposed approach the relevant boundary $S$ is connected from
the outset (see
Figure~\ref{fig:boundaries}.b). There is no need to refer to temporal
distances ``between'' boundaries. Temporal (or spatial)
distances related to a measurement process
can be evaluated on paths on the boundary using the
intrinsic metric only.

\section{Locality}
Apart from an analysis of the measurement process there are other
reasons to look for a TQFT type description of quantum gravity using
compact connected boundaries. An important reason 
is \emph{locality}.

Compact connected boundaries allow for an adaption of the mathematical
description to the size of the system considered. There is no need a
priori to include things (even empty space) at infinity. Of course,
going to infinity might not change the mathematical description much
or might even simplify it (e.g.\ asymptotic states in quantum field
theory). However, while this is certainly true in quantum mechanics
and quantum field theory on Minkowski space, it is very unlikely to be
true in a non-perturbative theory of quantum gravity.

Let me also mention that the program of Euclidean quantum gravity uses
compact manifolds and connected boundaries. However,
there the motivation is a mathematical one rather than a physical
one. Indeed the interpretation of corresponding quantities in that
program markedly differs from the interpretation that emerges in the
present context. For example, there one can define a ``wave function
of the universe'' \cite{HaHa:waveuniv}.
Here the interpretation of
essentially the same mathematical quantity
would be as giving rise to a functional on states that yields the amplitude
for a local measurement process on
the boundary of a finite region of space-time. This will be elaborated
elsewhere.

Talking about states or wave functions ``of the universe'' is often
motivated by ``realist'' interpretations of quantum mechanics, such as
the many-worlds interpretation (as clearly expressed in
\cite{Dew:qgrav1} for example). What I mean by this (without being
precise) is that one assigns a physical reality to
the wave function independent of any measurement process. One might
argue that such a point of view has no support from our
experimental evidence on quantum mechanics. I do not wish to take sides
in this debate here but emphasize that the local point of
view put forward here requires no particular interpretation of quantum
mechanics to be adopted a priori.

\section{Topological versus topological}
There is a folklore saying that classical general relativity is not a
topological theory and hence its quantization cannot be a topological
quantum field theory. However, this is in my view due to a misuse of
the word ``topological''. In the first instance it refers to the
classical
theory having no local degrees of freedom, in the second to the fact
that the background structures of the quantum theory are topological
manifolds and their
cobordisms (although ``differentiable'' would be more appropriate
here). The second does not imply the first. Indeed, consider the
classical limit. Then states on boundaries as in
Figure~\ref{fig:boundaries} which are ``peaked'' at a classical
metric, determine up to
diffeomorphisms essentially a unique classical solution of general
relativity ``inside''. There is no reason to think that a quantum
theory cannot incorporate this, for example as the dominant
contribution to a path integral. Indeed, this is for example a vital
ingredient of the Euclidean quantum gravity program.

One should not be misled by the fact that many interesting TQFTs that
have been constructed can be viewed as quantizations of topological
theories (e.g.\ \cite{Wit:qftjones}).
This seems rather related to the fact that the vector spaces associated
with boundaries there are finite dimensional (which would not be
expected for quantum gravity).

\section{Conclusions and Outlook}
In the non-relativistic context of the founding days of quantum
mechanics the picture of initial state, evolution and final state was
quite appropriate. I have argued that it becomes inadequate when
including space and time in the quantum mechanical realm itself.
Of course, I am not the first to express such a sentiment. However, I
have based my argument on the very principles of the quantum
mechanical measurement process itself. Furthermore, I have proposed a
way out, namely by adopting a TQFT type description for a quantum
theory of general relativity. Links between TQFT and quantum
general relativity have been suggested before (e.g.\ \cite{Cra:tqftqg}),
even to the extent of proposing a TQFT formulation \cite{Bar:qgqft}. The
physical interpretations, however, were rather different from the one
put forward here.
This is also the case for
other previous physical applications of TQFT which generally
maintain space-like boundaries.

This proposal has a number of implications. Among them are a somewhat
radical departure from the ``static'' Hilbert space picture of quantum
mechanics. Related to that is a necessary
duality between ``in'' and ``out'' states, or preparation and
observation. In particular, states are physically
meaningful even if
associated with boundaries that have time-like components. For
the interpretation of quantum mechanics conclusions might be drawn, in
particular the necessary ``delocalization'' in time of the ``collapse
of the wave function''. It also implies a shift in the interpretation of
quantities in present approaches to quantum gravity (e.g.\ the
``wave function of the universe'').

At the moment what I have presented is a proposal only, and many
crucial details are missing. It stands and
falls with the feasibility of formulating quantum theory in a
``general boundary'' way. Although designed for a quantum theory of
general relativity this can be adapted also to ordinary quantum
mechanics and quantum field theory. Then, the background structure is
not that of topological (or differentiable) manifolds but metric
manifolds, i.e.\ one would have a ``metric quantum field
theory''. At first sight the definition of ``particle'' seems to be a
major obstacle.
Surprisingly, it
appears that such a formulation is nevertheless possible (at least for
standard cases) for both quantum
mechanics and perturbative quantum field theory, and in a rather
uniform way. This will be demonstrated in forthcoming publications
\cite{Oe:boundary}.

Note that the present proposal also goes some way to suggest how to
modify present approaches to quantum gravity, with a
view towards obtaining physically meaningful amplitudes. This applies
notably to loop quantum gravity \cite{Rov:lqg} and spin foam
models \cite{Ori:spinfoamrev,Per:sfmodels}. A first step would be the
introduction
there of boundaries with both space- and time-like components. The
interpretation of amplitudes should then become clearer once the
quantum mechanics and quantum field theory situations have been worked
out.

\subsection*{Acknowledgements}
I would like to thank Carlo Rovelli for stimulating discussions,
Thomas Sch\"ucker for a critical reading of the manuscript and Domenico
Giulini for providing me with a reference. This
work was supported through a Marie Curie Fellowship of the European Union.

\bibliography{stdrefs}
\bibliographystyle{amsordx}

\end{document}